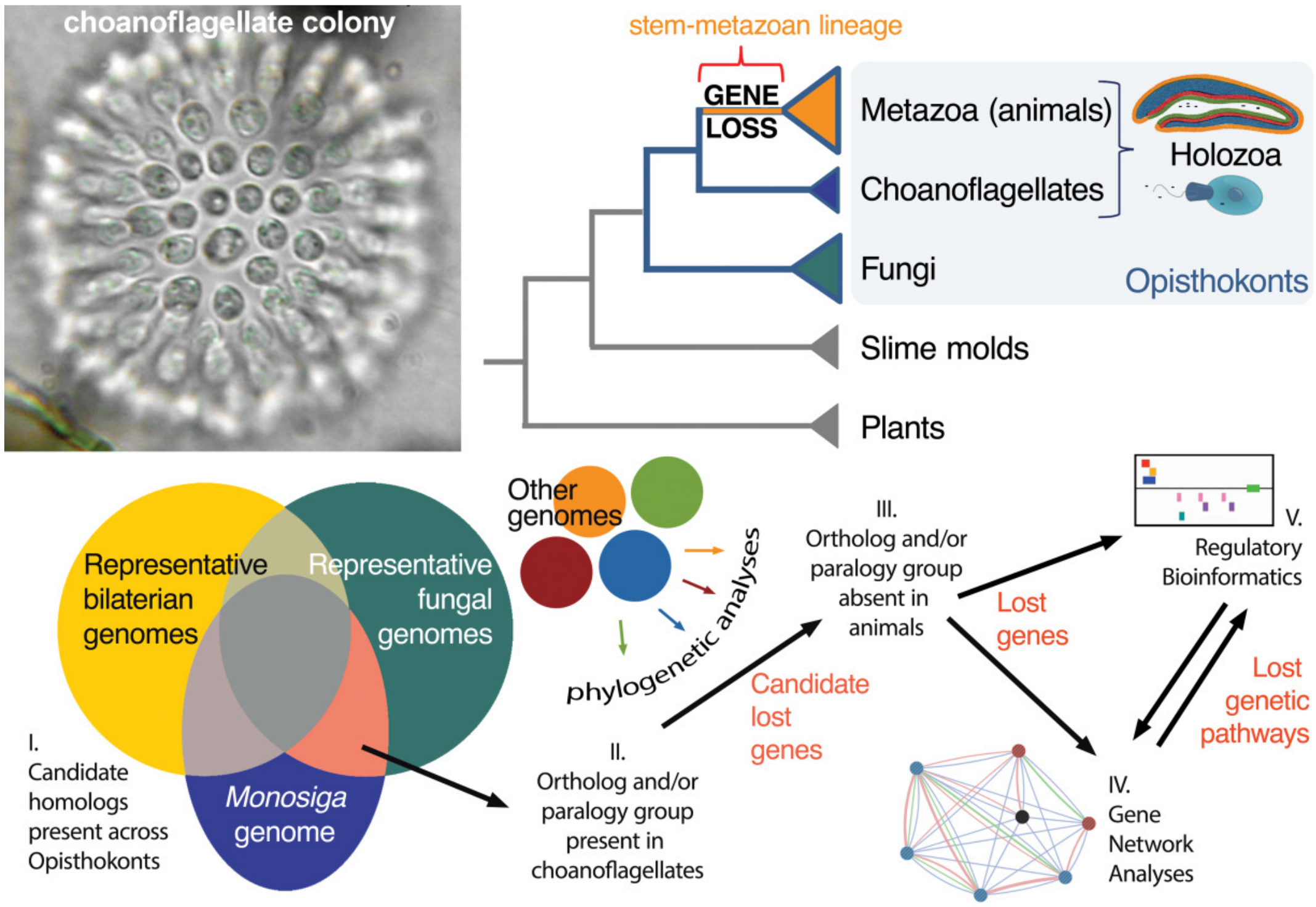

# Metabolic and Chaperone Gene Loss Marks the Origin of Animals: Evidence for Hsp104 and Hsp78 Sharing Mitochondrial Clients

Albert J. Erives (albert-erives@uiowa.edu) & Jan S. Fassler (jan-fassler@uiowa.edu)
Department of Biology, University of Iowa, Iowa City, IA, 52242-1324, USA

**Summary:** The evolution of animals involved acquisition of an emergent gene repertoire for gastrulation. Whether loss of genes also co-evolved with this developmental reprogramming has not yet been addressed. Here, we identify twenty-four genetic functions that are retained in fungi and choanoflagellates but undetectable in animals. These lost genes encode: (*i*) sixteen distinct biosynthetic functions; (*ii*) the two ancestral eukaryotic ClpB disaggregases, Hsp78 and Hsp104, which function in the mitochondria and cytosol, respectively; and (*iii*) six other assorted functions. We present computational and experimental data that are consistent with a joint function for the differentially localized ClpB disaggregases, and with the possibility of a shared client/chaperone relationship between the mitochondrial Fe/S homoaconitase encoded by the lost *LYS4* gene and the two ClpBs. Our analyses lead to the hypothesis that the evolution of gastrulation-based multicellularity in animals led to efficient extraction of nutrients from dietary sources, loss of natural selection for maintenance of energetically expensive biosynthetic pathways, and subsequent loss of their attendant ClpB chaperones.

## Introduction

By the time of the last common ancestor of animals, the uniquely animal process of gastrulation had evolved. Gastrulation, while diverse in form, is a highly regulated re-arrangement and migration of cells producing a layered multicellular body with an internalized "endoderm"-derived digestion system. Many post-genomic studies have detailed the evolutionary expansion of eukaryotic transcription factor (TF) families and the emergence of new developmental genes during this major evolutionary transition [1-5]; these genes include those controlling the gastrulation program, which is shared from humans to sponges [6,7]. However, identification of pre-metazoan gene losses may be equally important because they may indicate the context, origins, and constraints associated with new transcription programs.

No study has yet focused on systematically identifying genes that are widely retained in the closest relatives of animals but undetectable in animals themselves. Such studies require resources and methodologies that have not been available or mature enough until recently. These include: (i) relevant genome assemblies; (ii) robust, accurate, orthology-calling pipelines based on reliable gene annotations; and (iii) powerful gene association engines and data sets from model genetic systems that would allow candidate losses to be placed into genetic pathways and regulatory programs. Now, assembled genomes are increasingly available from such early-branching animal lineages as sponges (*e.g.*, *Amphimedon queenslandica* [4]), ctenophores (*Mnemiopsis leidyi* [8]), placozoans (*Trichoplax adhaerens* [9]), as well diverse bilaterians, including chordate and non-chordate deuterostomes and ecdysozoan and lophotrochozoan protostomes [10]. Also available are genomes of animal's closest eukaryotic relatives, including its sister clade of choanoflagellates (*Monosiga brevicollis* [11] and *Salpingoeca rosetta* [12]), as well as filastereans (*e.g.*, *Capsaspora owczarzaki* [13]), and diverse fungi from ascomycetes, basidiomycetes, and chytridiomycetes (*e.g.*, *Batrachochytrium dendrobatidis*). In addition, Ensembl's Maximum-Likelihood based orthology-calling pipeline EnsemblCompara [14] and BioMart query engine [15] can be used to identify candidate lost genes, which can then be investigated and confirmed with formal phylogenetic analysis on a gene by gene basis. Finally, powerful algorithms have been designed to query genetic, transcriptomic, proteomic, and sub-cellular localization data sets and to measure meaningful enrichment of functional categories [16,17]. Such association algorithms require large amounts of data and data types for sufficient power. Arguably, the most comprehensive data is available for the yeast, Saccharomyces cerevisiae. For this reason, this organism is the favored system for understanding at a quantitative and analytical level how genome-encoded functions interact to produce complex phenotypes [18,19].

Here we use comparative and integrative genomic approaches to identify and analyze a set of eukaryotic genes that are retained and detectable in the closest-living relatives of animals, but are no longer detectable in animals themselves. Interpreted parsimoniously, these undetectable genes correspond to genetic functions lost in the stem-metazoan lineage, which begins just *after* the latest common ancestor of animals and choanoflagellates and ends just *before* the latest common ancestor of animals. We find that the majority of these genes correspond to lost biosynthetic pathways, including biosynthesis of L-histidine, L-lysine, riboflavin, and L-tryptophan. Intriguingly, we find specific connections between the two lost lysine biosynthetic genes, *LYS4* and *LYS20*, and the lost eukaryotic ClpB chaperones encoded by *HSP78* and *HSP104*, which have roles in protein folding in the mitochondria and cytoplasm, respectively. Computational and experimental data suggest that the two ClpB chaperones may work together. In this study we investigate the hypothesis that, while Hsp78 and Hsp104 are differentially compartmentalized, they may share a subset of clients in nuclear-encoded, cytoplasmically-translated, mitochondrially-imported proteins. Tandem affinity purification (TAP) of chaperones and identification of interacting proteins with mass spectrometry (MS) (TAP-MS) data in yeast reveal that the Hsp104 and Hsp78 interact with Lys4 and Lys20 but not with the proteins encoded by the other lost genes. Taken together our results are consistent with the hypothesis that Lys4 and Lys20 are prone to misfolding and require ClpB chaperones after thermal stress and suggest that some of the genes identified in our study do not necessarily represent independent losses but may instead reflect a series of dependent events. Our genomic approach and results are likely to improve our understanding of evolutionary origins and unique nature of animal metabolism.

## Results

### Identification of twenty-five genes retained in choanoflagellates but not animals.

To identify protein-coding genes retained in the closest eukaryotic relatives of animals but which are not otherwise detectable in animals (henceforth the "lost genes"), we made two methodological decisions. First, we chose *Saccharomyces cerevisiae* as a reference genome despite the possibility that we would not identify some lost genes associated with derived traits in the budding yeast. However, by using the budding yeast we avail ourselves of the large number of existing genetic, transcriptomic, proteomic, and sub-cellular localization data sets. Second, to avoid false-positive gene losses due to deficiencies in assembled genomes, we sampled diverse non-animal genomes without over-sampling any particular taxon, and focused on the fewest genomes possible without sacrificing taxonomic diversity. For Dikarya fungi we worked with the ascomycetes, *S. cerevisiae* and *Schizosaccharomyces pombe* genomes; for the basidiomycetes, *Cryptococcus neoformans* and *Puccinia graminis* (Ensembl BioMart); and for chytridiomycetes, *Batrachochytrium dendrobatidis* (JGI). For Filasterea, we chose the *Capsaspora owczarzaki* (Broad Institute) genome [13]. Last, for choanoflagellates we used *Monosiga brevicollis* (Joint Genome Institute, JGI) and the colonial species *Salpingoeca rosetta* (Broad Institute) [11,20].

To generate a manageable candidate gene list we used Ensembl's BioMart query engine [14,15,21] to iteratively identify orthologs that are present in the four listed genomes of Dikarya (Fig. 1) but absent in the two bilaterians, *Drosophila melanogaster* and *Homo sapiens*, each of which has an excellent genome assembly with few gaps. We also accepted lineage-specific duplications of a single shared ancestral gene to avoid losing bona fide candidates arising from whole genome duplications in any one lineage. Of the 6,692 protein-coding genes in *S. cerevisiae* (Ensembl R64-1-1), 802 genes survived this initial screen. We then queried the Joint Genome Institute's (JGI) BLASTP server (genome.jgi.doe.gov) with each yeast protein to identify those for which the predicted protein-coding sequences from the choanoflagellate *M. brevicollis* (Monbr1_best_proteins) matched the yeast proteins with a relatively stringent BLASTP expect value threshold (E-value lower (better) than 1.0E-30). This reduced our list to 210 yeast proteins. We used a threshold E-value of 1.0E-10 to remove yeast proteins that either had no matches or were not detectable for the colonial choanoflagellate *S. rosetta* and the filasterean *C. owczarzaki*. We also removed those that had E-values in animals smaller (*i.e.*, better) than any of the three listed non-animal holozoans (the choanoflagellates and filasterean) (Fig. 1). For animals, we used the sponge *Amphimedon queenslandica* (NCBI) [4], the ctenophore *Mnemiopsis leidyi* (NHGRI) [8], the placozoan *Trichoplax adhaerens* (NCBI) [9], and the lophotrochozoans *Lottia gigantea*, *Capitella teleta*, and *Helobdella robusta*, all of which have complete genome sequence assemblies [10]. With these genomic comparisons, we find that there are twenty-seven yeast genes that are largely retained outside of animals in the genomes we analyzed (Fig. 1, and Table S1). Six of these genes (*TRR1*/*TRR2*, *LYS20*/*LYS21*, and *SCT1*/*GPT2*) correspond to lineage-specific duplications in yeast (connected genes in Fig. 1). Thus, we find twenty-four genetic functions that are definitively lost in animals or at least not retained at the same level of conservation as they are in non-metazoans.

A heat map of the E-values for the best BLASTP hits in each of the genomes shows that most "lost" candidate genes are undetectable in metazoans, but that a few have matches with poor E-values relative to the non-metazoans (Fig. 1B, Table S1), or else could be shown to be environmental contaminants or horizontal transfers from distinct taxa (Fig. 1B, taxonomic names given). Some were found to be non-orthologous sequences that are members of other gene clades in a superfamily defined by a common domain (Fig. 1B, gene names given). In only a single case (*HIS2*) was there a weak match from an animal genome (*Trichoplax adhaerens*) that was most closely-related to the orthologous gene from choanoflagellates or the filasterean *Capsaspora*. The single possible hit in *Nematostella vectensis* for *HIS7* (Fig. 1) corresponds to the proteobacterial version of this gene and not a gene related to opisthokont or eukaryotic *HIS7*, and similar results were found for *TRP5* and others as indicated (Fig. 1B). One gene labeled "clpB" in many animals and which matches yeast Hsp104 and Hsp78, is actually a fusion gene between an N-terminal ankyrin domain and a C-terminal partial sequence of a bacterial *clp* gene (Fig. S1). Phylogenetic analysis of this "*ANKCLP*" fusion gene (Fig. S1) indicates that it arose in the stem-holozoan lineage such that it still occurs alongside the *HSP104* and *HSP78* genes of both choanoflagellates.

We found that the closest matches in animal sequences to either Hsp104 or Hsp78 are partial fragments of bacterial ClpB in *Amphimedon* (Fig. 1) or the unrelated bacterial ClpC in *Nematostella* (Fig. S1). We used Bayesian Inference and sampled mixed amino acid substitution models to infer a ClpB/Hsp78/Hsp104 phylogenetic tree consisting of the fungal and holozoan sequences as well as the best matches to the 456 amino acid *Amphimedon* sequence, which is only half as long as the proteins in the eukaryotic and bacterial ClpB clades (Fig. 2). Our results definitively place the predicted partial peptide sequence from *Amphimedon* in a ClpB clade alongside orthologs from $\gamma$-proteobacteria (Fig. 2). The sequences most closely related to the *Amphimedon* fragment are from bacterial endosymbionts of marine tube worms (Fig. 2). This suggests that this sequence is either an environmental contaminant or a partial horizontal gene transfer from endosymbiotic bacteria.

The second best match to Hsp104 in metazoans was an even shorter peptide sequence of 404 amino acids from *Nematostella vectensis* (Fig. S1). This predicted peptide sequence most closely matches ClpC peptides from flavobacteria when the entire database is queried. Queries of Hsp104 sequences to the taxonomically unrestricted animal database reveal other spurious hits, which can be discounted for similar reasons (Fig. S1B). Thus, these results suggest that animals truly lost both Hsp104 and Hsp78.

Non-orthology for some weak matches in animal genomes is seen for the gene *LYS4*, which encodes mitochondrial homoaconitase. The closest match in animals to *LYS4* is the *ACO1* gene, which encodes the cytoplasmic aconitase of the tricarboxylic acid cycle and is broadly retained in both animals and non-animals as expected (Fig. 3A, Bayesian Inference; and Fig. 3B, Neighbor-Joining). For most lost genes, there are neither orthologs nor distantly related homologs in animals even when such genes are present in bacteria (*e.g.*, Fig. 3C, *LYS20*).

To summarize the results of our gene loss approach, we find there are at least twenty-four highly retained genetic functions that are specifically lost in animals. To determine the extent to which this query strategy for lost animal genes is biased by the requirement for presence in fungi, we performed a second independent query focused on gene families not present in *Saccharomyces cerevisiae*, a key species in our first analysis. We identified 14 PFAM families present in the social slime mold *Dictyostelium discoideum* and the choanoflagellate *Monosiga brevicollis*, but not in animals (initially *Homo sapiens*, *Drosophila melanogaster*, *Nematostella vectensis*, *Trichoplax adhaerens*, and *Amphimedon queenslandica*) and the yeast *S. cerevisiae*. We used the *Dictyostelium* predicted protein sequences (19 peptide sequences) corresponding to these 14 PFAM domains as queries in a BLAST analysis to identify homologs in choanoflagellates. The *Dictyostelium* gene *PURB* (PF08238), which encodes an adenylosuccinate lyase (467 amino acids), is present in *Monosiga brevicollis* (2e-129), *Salpingoeca rosetta* (1e-134), and *Capsaspora owczarkii* (8e-169), but NOT in fungi or animals. A second *Dictyostelium* gene *PHAZ* (PF10503), which encodes a putative polyhydroxybutyrate depolymerase (341 amino acids), is present as two copies (paralogs) in each choanoflagellate genome (E-values range from 1e-87 to 4e-76), but we found this to have matches in the non-eumetazoan animal lineages. In summary, this alternate strategy identified six lost animal genes (*GDH2*, *HSP104*, *HSP78*, *PAN6*, *PURB*, *TRR1/2*) five of which were identified by the fungi-based screen (see Fig. 1A, right-hand column). In total, we found twenty-five distinct gene functions maintained in choanoflagellates (and in fungi or slime mold) but definitively lost in animals (File S1).

**Fig. 1. Twenty-five genetic functions undetectable in animals.**

(**A**) Heat map of twenty-seven genes (first twenty-seven rows with gene names) from the budding yeast *S. cerevisiae* (*S. cer.*) and their orthologs in related organisms with whole genome sequence assemblies and their closest matches in animals. Six of the twenty-seven yeast genes are the result of lineage-specific duplications in yeast (indicated by dark gray brackets, left of table), leaving twenty-four genetic functions either lost or not measurably conserved in metazoans. The heat map is based on the BLASTP expect values of the best match to the indicated yeast gene (see key, lower right of table). Species abbreviations: *S. cer.*, *Saccharomyces cerevisiae*; *S. pom.*, *Schizosaccharomyces pombe*; *C. neo.*, *Cryptococcus neoformans*; *P. gra.*, *Puccinia graminis*; *B. den.*, *Batrachochytrium dendrobatidis; C. owc.*, *Capsaspora owczarzaki; M. bre.*, *Monosiga brevicollis*; *S. ros.*, *Salpingoeca rosetta; A. que.*, *Amphimedon queenslandica*; *T. adh.*, *Trichoplax adhaerens; M. lei., Mnemiopsis leidyi; N. vec., Nematostella vectensis; L. gig., Lottia gigantea; C. tel.*, *Capitella teleta*; *H. rob.*, *Helobdella robusta*; *D. mel.*, *Drosophila melanogaster*; and *H. sap.*, *Homo sapiens.* Other abbreviations: Chytridio., Chytridiomycota; Placo., Placozoa; Ctenoph., Ctenophora; Ecdyso., Ecdysozoa; and Chor., Chordata. The last row is from an alternative screen for lost animal genes (see text) in which Amoebozoa PFAM domains that are absent in animals were identified (five genes already identified plus one new gene, *PURB*).

(**B**) Close-up of the lost animal gene columns from panel A. This table shows either the taxonomic origin of a weak animal match, or the name of the gene if it is not directly related (*i.e.,* not orthologous) to the named yeast gene (first column) for all matches with BLASTP expectation values < 1e-20. Most of the animal matches to the candidate lost genes correspond to lineage-specific horizontal gene transfers and/or environmental contaminants from unrelated (non-opisthokont) clades. Some matches correspond to non-orthologous genes such as *ANKCLP* (Fig. S2) in the case of the eukaryotic clpB genes *HSP78* and *HSP104*, or *ACO1* in the case of *LYS4* (Fig. 3). Only *HIS2* might have been lost soon after early animal radiation based on the presence of a weak match in *Trichoplax adhaerens*. This gene is most similar to the version from *Capsaspora* and not to either of the two choanoflagellates, so it is of uncertain origin.

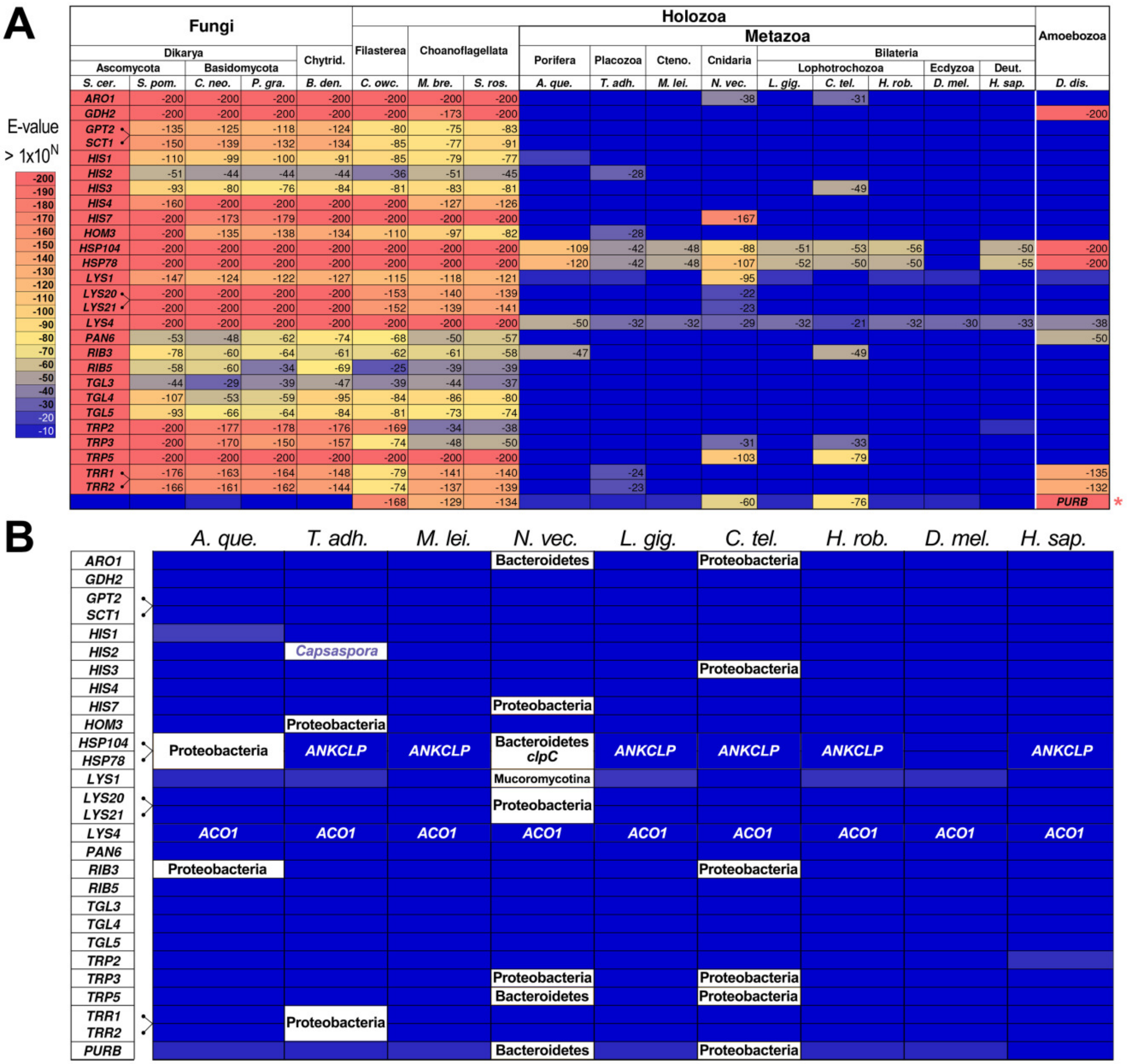
A
Fungi
Holozoa
Metazoa
Amoebozoa
Dikarya
Ascomycota
Basidiomycota
Chytrid.
Filasterea
Choanoflagellata
Porifera
Placozoa
Cteno.
Cnidaria
Bilateria
Lophotrochozoa
Ecdyzoa
Deut.
S. cer. S. pom. C. neo. P. gra. B. den. C. owc. M. bre. S. ros. A. que. T. adh. M. lei. N. vec. L. gig. C. tel. H. rob. D. mel. H. sap. D. dis.
E-value
> 1x10^N
-200 -190 -180 -170 -160 -150 -140 -130 -120 -110 -100 -90 -80 -70 -60 -50 -40 -30 -20 -10
ARO1 -200 -200 -200 -200 -200 -200 -200 -38 -31
GDH2 -200 -200 -200 -200 -200 -173 -200 -200
GPT2 -135 -125 -118 -124 -80 -75 -83
SCT1 -150 -139 -132 -134 -85 -77 -91
HIS1 -110 -99 -100 -91 -85 -79 -77
HIS2 -51 -44 -44 -44 -36 -51 -45 -28
HIS3 -93 -80 -76 -84 -81 -83 -81 -49
HIS4 -160 -200 -200 -200 -200 -127 -126
HIS7 -200 -173 -179 -200 -200 -200 -200 -167
HOM3 -200 -135 -138 -134 -110 -97 -82 -28
HSP104 -200 -200 -200 -200 -200 -200 -200 -109 -42 -48 -88 -51 -53 -56 -50 -200
HSP78 -200 -200 -200 -200 -200 -200 -200 -120 -42 -48 -107 -52 -50 -50 -55 -200
LYS1 -147 -124 -122 -127 -115 -118 -121 -95
LYS20 -200 -200 -200 -200 -153 -140 -139 -22
LYS21 -200 -200 -200 -200 -152 -139 -141 -23
LYS4 -200 -200 -200 -200 -200 -200 -200 -50 -32 -32 -29 -32 -21 -32 -30 -33 -38
PAN6 -53 -48 -62 -74 -68 -50 -57 -50
RIB3 -78 -60 -64 -61 -62 -61 -58 -47 -49
RIB5 -58 -60 -34 -69 -25 -39 -39
TGL3 -44 -29 -39 -47 -39 -44 -37
TGL4 -107 -53 -59 -95 -84 -86 -80
TGL5 -93 -66 -64 -84 -81 -73 -74
TRP2 -200 -177 -178 -176 -169 -34 -38
TRP3 -200 -170 -150 -157 -74 -48 -50 -31 -33
TRP5 -200 -200 -200 -200 -200 -200 -200 -103 -79
TRR1 -176 -163 -164 -148 -79 -141 -140 -24 -135
TRR2 -166 -161 -162 -144 -74 -137 -139 -23 -132
-168 -129 -134 -60 -76 PURB *
B
A. que. T. adh. M. lei. N. vec. L. gig. C. tel. H. rob. D. mel. H. sap.
ARO1 GDH2 GPT2 SCT1 HIS1 HIS2 HIS3 HIS4 HIS7 HOM3 HSP104 HSP78 LYS1 LYS20 LYS21 LYS4 PAN6 RIB3 RIB5 TGL3 TGL4 TGL5 TRP2 TRP3 TRP5 TRR1 TRR2 PURB
Bacteroidetes
Proteobacteria
Capsaspora
Proteobacteria
Proteobacteria
Proteobacteria
Proteobacteria
ANKCLP
ANKCLP
Bacteroidetes clpC
ANKCLP
ANKCLP
ANKCLP
ANKCLP
Mucoromycotina
Proteobacteria
ACO1 ACO1 ACO1 ACO1 ACO1 ACO1 ACO1 ACO1 ACO1
Proteobacteria
Proteobacteria
Proteobacteria
Proteobacteria
Bacteroidetes
Proteobacteria
Proteobacteria
Bacteroidetes
Proteobacteria

**Fig. 2. Phylogenetic analyses confirming absence of intact *clpB* genes in animals.**

(**A**) Phylogenetic inference using Bayesian MCMC methods shows that the closest matching sequence in known animal genome sequences is a partial ClpB fragment in the sponge *Amphimedon queenslandica*. This fragment does not contain both nucleotide binding domains (NBDs) seen in bacterial or eukaryotic (Hsp78 and Hsp104) ClpB sequences, which are at least 800 amino acids long. The top three sequence hits in GenBank databases using the *Amphimedon* ClpB fragment as query are included. Two of these are intact ClpB protein sequences from bacterial (γ-proteobacteria) endosymbionts of marine tubeworms. Topological convergence was achieved after 300,000 generations with a relative burn-in phase of 25% (MrBayes) [80-82]. Mixed models were tested, but WAG [83] was the sole model favored upon completion. Average standard deviation of split frequencies was $< 0.002$.The tree was rooted between the bacterial clpB clade, including mitochondrial Hsp78, and the eukaryotic Hsp104 clade.

(**B**) Maximum likelihood phylogenetic analysis of the ClpB family with choanoflagellates and chytridiomycetes (*Batrachochytrium dendrobatidis* and *Gonapodya prolifera*) representing fungal lineages shows that Hsp78 is the mitochondrially inherited *clpB* gene based on its sister grouping with α-proteobacteria. Dikarya sequences were omitted to simplify the tree. Numbers indicate bootstrap support from 500 replicates using MEGA5 [85]. (**C**) A simple, unrooted, neighbor-joining tree of the highly-conserved *clpB* genes across all of life shows that the ClpB functions are maintained in all major domains of life and in all endosymbiotic eukaryotic compartments. The ubiquity of the *clpB* genes makes the loss in animals more striking.

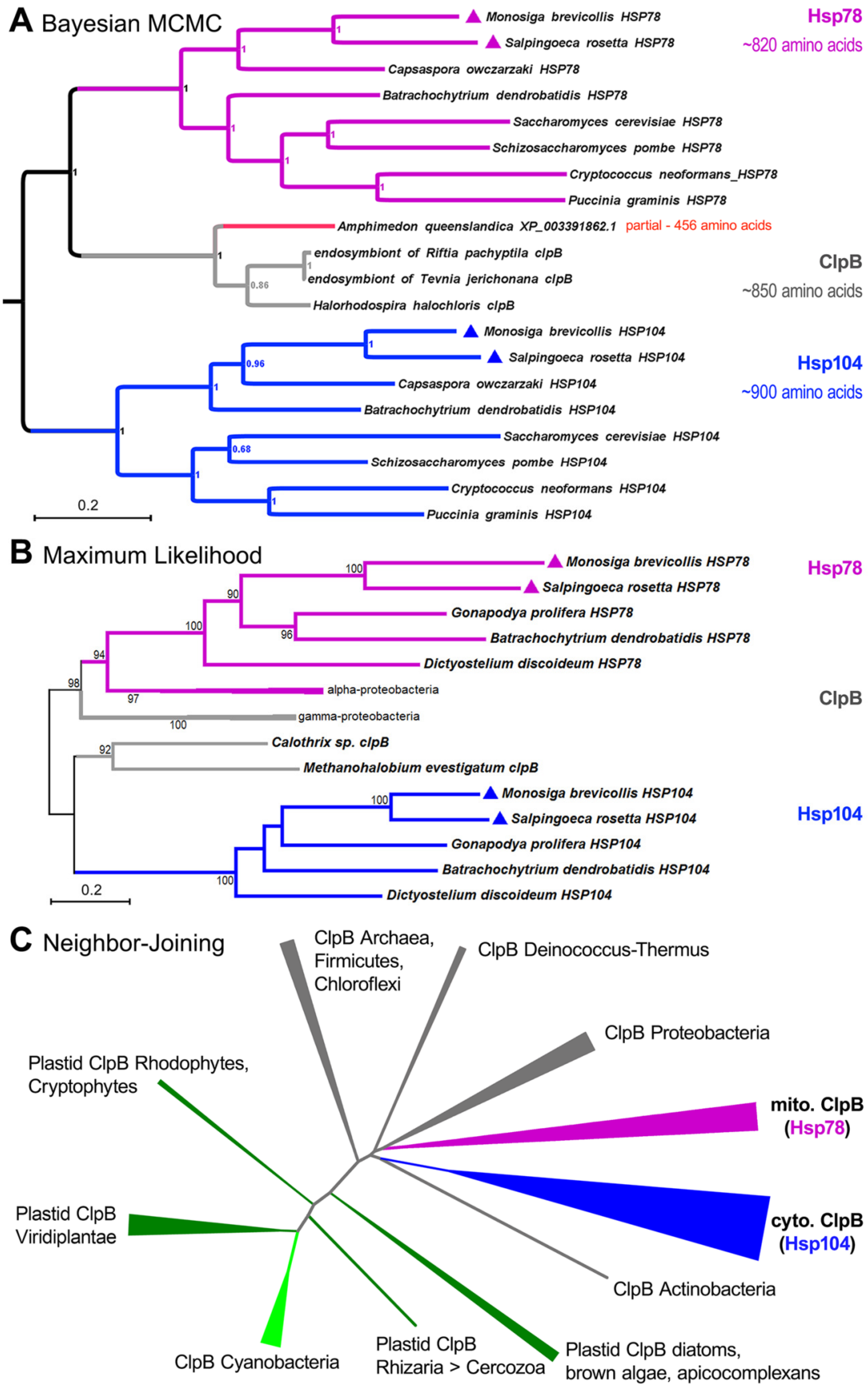

**Fig. 3. Phylogenetic analysis confirming absence of *LYS4* and *LYS20* in animals.**

(**A**) Phylogenetic inference using Bayesian MCMC methods shows that the closest sequence matches in animal genome sequences to the *LYS4* genes are the *ACO1* genes, which encodes the aconitase enzyme of the tricarboxylic acid (TCA) cycle. Choanoflagellates (triangles) have *LYS4* and *ACO1* but animals (squares) have only *ACO1*. The tree is rooted between the two gene clades for Lys4 and Aco1.

(**B**) A representative neighbor-joining (NJ) tree shows the presence of *LYS4* (blue) and *ACO2* (pink) in choanoflagellates (triangles), but not in animals (squares), for which sequences could not be found. One related gene *ACO1* (maroon), which encodes an enzymatic activity of the TCA cycle, is still present in animals, while *LEU1* (mustard yellow), a second related gene, involved in the leucine biosynthesis as part of the branched chain amino acid pathway, is lost in both choanoflagellates and animals. *ACO2* could not be found in the colonial choanoflagellate, *Salpingoeca rosetta*. Numbers indicate bootstrap support from 500 replicates. The tree is rooted between the Leu1 outgroup clade and theLys4/Aco1/Aco2 superclade.

(**C**) A representative neighbor-joining tree shows the presence of *LYS20*/21 gene in choanoflagellates (triangles) and other organisms. *LYS20* and *LYS21* correspond to yeast specific duplications of this genetic function. No related homologs are found in animals (Fig. 1). Distantly-related bacterial outgroups do however possess these genes. The tree is rooted between the eukaryotic Lys20/21 clade and the bacterial (Deinococcus-Thermus) gene clade.

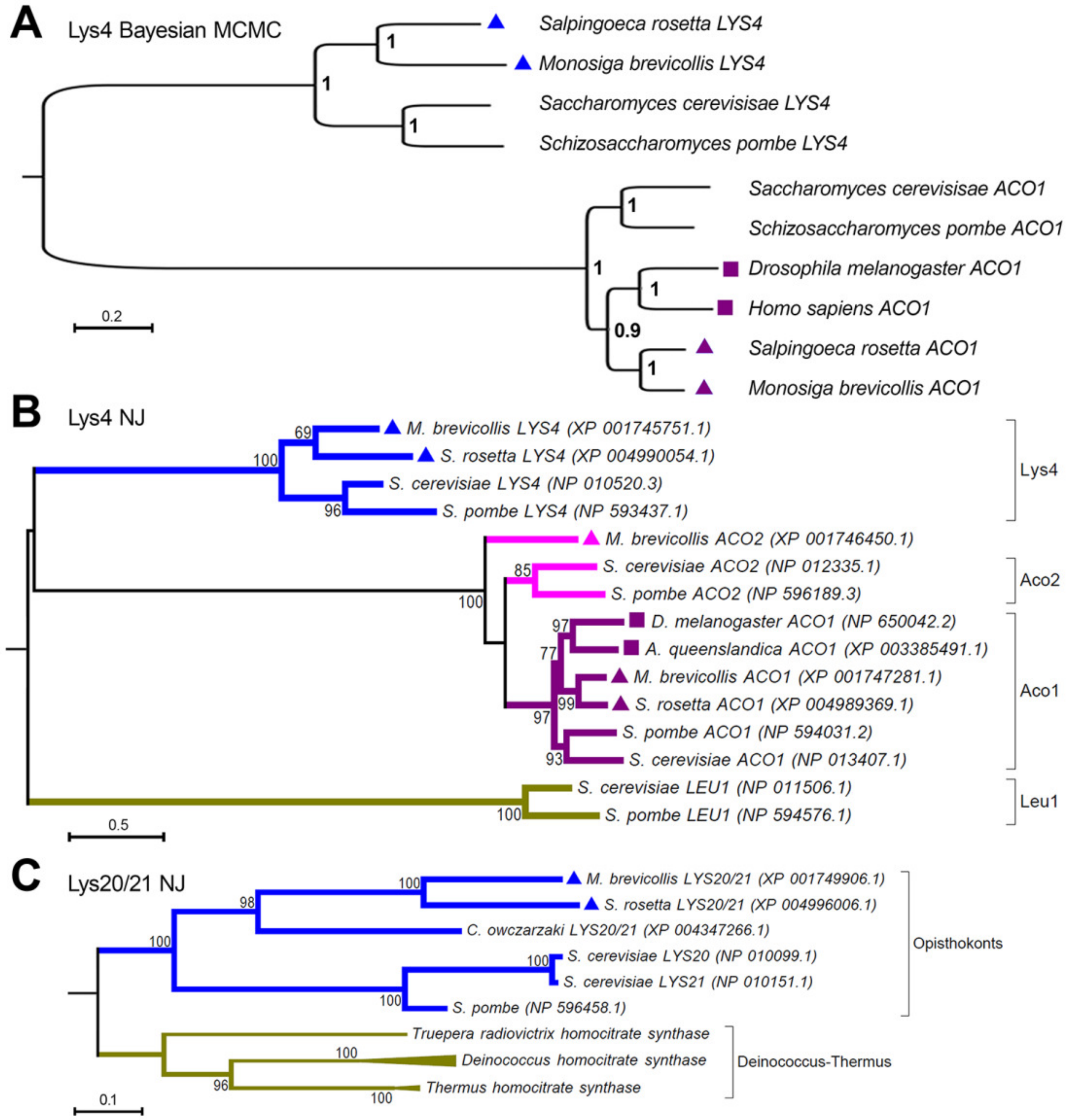
A
Lys4 Bayesian MCMC
Salpingoeca rosetta LYS4
Monosiga brevicollis LYS4
Saccharomyces cerevisisae LYS4
Schizosaccharomyces pombe LYS4
Saccharomyces cerevisisae ACO1
Schizosaccharomyces pombe ACO1
Drosophila melanogaster ACO1
Homo sapiens ACO1
Salpingoeca rosetta ACO1
Monosiga brevicollis ACO1
0.2
0.9
B
Lys4 NJ
M. brevicollis LYS4 (XP 001745751.1)
S. rosetta LYS4 (XP 004990054.1)
S. cerevisiae LYS4 (NP 010520.3)
S. pombe LYS4 (NP 593437.1)
M. brevicollis ACO2 (XP 001746450.1)
S. cerevisiae ACO2 (NP 012335.1)
S. pombe ACO2 (NP 596189.3)
D. melanogaster ACO1 (NP 650042.2)
A. queenslandica ACO1 (XP 003385491.1)
M. brevicollis ACO1 (XP 001747281.1)
S. rosetta ACO1 (XP 004989369.1)
S. pombe ACO1 (NP 594031.2)
S. cerevisiae ACO1 (NP 013407.1)
S. cerevisiae LEU1 (NP 011506.1)
S. pombe LEU1 (NP 594576.1)
Lys4
Aco2
Aco1
Leu1
0.5
C
Lys20/21 NJ
M. brevicollis LYS20/21 (XP 001749906.1)
S. rosetta LYS20/21 (XP 004996006.1)
C. owczarzaki LYS20/21 (XP 004347266.1)
S. cerevisiae LYS20 (NP 010099.1)
S. cerevisiae LYS21 (NP 010151.1)
S. pombe (NP 596458.1)
Truepera radiovictrix homocitrate synthase
Deinococcus homocitrate synthase
Thermus homocitrate synthase
Opisthokonts
Deinococcus-Thermus
0.1

## Animal-specific co-loss of the ClpB chaperones, Hsp78 and Hsp104.

Both *HSP104* and *HSP78* are nuclear encoded eukaryotic orthologs of bacterial *clpB*, and encode AAA+ (ATPases Associated with diverse cellular Activities) family ATPases with two P-loop nucleotide-binding domains (NBDs). Together with an N terminal domain, the two NBDs form a three tier hexameric ring with a central pore through which misfolded, aggregated peptides are translocated prior to refolding [22,23]. Hsp78 localizes and functions in the mitochondria [24-26], while Hsp104 localizes to the cytoplasm and nucleus [27]. Because of Hsp104's role in amyloid/cross-β disaggregation in yeast [28], the reason for the loss of the highly-conserved gene *HSP104* in animals has been a source of much speculation [29].

In eukaryotes, the nuclear encoded *HSP78* gene was acquired from the mitochondrial ancestor [30]. Our phylogenetic analyses repeatedly confirm that choanoflagellates, like fungi, have both *HSP78* and *HSP104,* which correspond to the mitochondrial and eukaryotic *clpB* orthologs that are absent in animals (Fig. 2B; also see close relationship of *HSP78* clade to $\alpha$-proteobacterial *clpB*). The loss of these genes in the stem-metazoan lineage is more remarkable when considering that all organisms and organelles (e.g., eukaryotic mitochondrion, plant chlorplast), except those in animals, appear to retain ClpB functions, even when such genes are transferred to the nucleus (Fig. 2C).

## Table 1 | A perfect HSE4 element is specific to a *clpB* refolding regulon in *S. cerevisiae*.

| # | Model* | Model description* | Genomic distribution | Precision** |
|---|---|---|---|---|
| 1 | 2-way DP-I | TTTTCCAGAAT.TTCTAGAAG | **ClpB**: *HSP104*, *HSP78*. | 100% (2/2) |
| 2 | 2-way PWM (Ψ-count: 0.3) | t 55550000052550500000<br>c 00005500000100500000<br>g 00000005000100000500 5<br>a 00000050550100005055 0 | **ClpB**: *HSP104*, *HSP78*.<br>**Hsp70**: *SSA1*. **DnaJ**: *SIS1*, *APJ1*.<br>**Other**: *BTN2*, *HAP4*, *TMA19*, *ADD37*. | 50% (6/12) |
| 3 | 3-way PWM (Ψ-count: 0.2) | t 46660000004266060000 0<br>c 20006600002000600000 0<br>g 00000006000200000600 4<br>a 00000060660200006066 2 | **ClpB**: *HSP104*, *HSP78*.<br>**Hsp70**: *SSA1*. **DnaJ**: *SIS1*, *APJ1*.<br>**Other**: *BTN2*, *UBI4*, *ADD37*. | 78% (7/9) |
| 4 | HSE4 | ..TTC..GAA..TTC..GAA. | **ClpB**: *HSP104*, *HSP78*.<br>**Hsp70**: *SSA1*. **DnaJ**: *SIS1*.<br>**Other**: *BTN2*, *UBI4*, *HSP42*. | 78% (7/9) |
| 5 | Perfect HSE4 | ..TTCYRGAA..TTCYRGAA. | **ClpB**: *HSP104*, *HSP78*.<br>**Hsp70**: *SSA1*. **DnaJ**: *SIS1*.<br>**Other**: *BTN2*, *UBI4*. | 100% (6/6) |
| 6 | Skipped HSE4 + 3x STREs | ..TTC..GAA.......GAA.<br>+ 3x 5′-CCCCT | **ClpB**: *HSP104*, *HSP78*.<br>**Other**: *HSP42*. | 100% (3/3) |

Abbreviations: DP-I, dot plot motif I; IUPAC, International Union of Pure and Applied Chemistry; PWM, Position-Weighted Matrix; Ψ, pseudo-count correction. * Models 1–3 describe the genomic distributions of the *clpB*-specific HSE4 element using different descriptions. IUPAC DNA codes are used as needed except 'N' is represented by a dot ('.'). Y = pYrimidine (C or T), and R = puRine (A or G). ** Percentages refer to the fraction of all genomic loci matching the model signature that are known to be involved in prion homeostasis. All such loci are listed in bold in the penultimate column. Hits in CDS regions are not listed but are counted for the precision metric.

## Table 2 | The *clpB* genes are co-regulated by HSE4s in Fungi and Holozoa.

| HSE4 motif models* | | *Saccharomyces cerevisiae* | *Salpingoeca rosetta* |
|---|---|---|---|
| 1 | TTCYRGAA..TTCYRGAA | *HSP78*, *HSP104*, *SSA1*, *BTN2*, *UBI4*, *SIS1* | *HSP78* |
| 2 | TTC.RGAA..TTCYRSAA | *HSP78*, *HSP104*, *SSA1*, *BTN2*, *UBI4*, *SIS1*, +2 | *HSP78*, *HSP104*, *BMS1*, +1 in ORF |
| 3 | TTC..GAA..TTCYRSAA | *HSP78*, *HSP104*, *SSA1*, *BTN2*, *UBI4*, *SIS1*, +3 | *HSP78*, *HSP104*, *BMS1*, *UBA1*, +2, +1 ORF |
| 4 | TTC..GAA..TTC..GAA | *HSP78*, *HSP104*, *SSA1*, *BTN2*, *UBI4*, *SIS1*, +3 | *HSP78* |

* Each locus matching the indicated motif in each genome is either listed as a named gene or enumerated. For example, "+2" means "plus two more loci" with matches in upstream regulatory region and "+2" in ORF" means "plus two more in ORFs".

## *HSP104* and *HSP78* are co-regulated by a specialized *cis*-element, HSE4.

To explore the possibility that joint loss of the ClpB orthologs reflects a joint requirement and/or a joint function for the proteins, we examined their regulation and, as described below, we also assessed their double mutant phenotype. To test the presence of shared *clpB* regulatory signatures, we initially focused on *S. cerevisiae*, for which predictions of co-regulation could be confirmed independently by gene expression meta-analyses of the large number of published transcriptome studies in yeast [31]. Regulatory motifs common to the upstream regions of the *S. cerevisiae HSP78* and *HSP104* genes were identified using BLASTN-based dot plot analysis. A comparison of 320 bp regions upstream of the *HSP78* and *HSP104* open reading frames (Fig. 4A) revealed three serially-aligned blocks of sequence whose length is consistent with their function as possible protein binding sites (Fig. 4B) and that are found in orthologous loci in other *Saccharomyces* species (Fig. 4C, track 6). These analyses led to the identification of three shared signature features (Fig. 4C): (***i***) an Hsf1 binding Element characterized by a series of four inverted 5′-**nGAAn** repeats [32-36] shown in blue (HSE4, track 3), and in red and orange (DP-I = dot-plot motif 1, tracks 1 and 2); (***ii***) a triplet array of Stress-Response Element (STREs) that bind the paralogous zinc finger transcription factors, Msn2 and Msn4 [37,38]; and (***iii***) a core promoter element that resembles a TATA-box with additional sequences immediately downstream (DP-III, track 5). Binding of Hsf1 to the HSE and binding of Msn2, and Msn4 to the STRE have been observed under oxidative stress, heat shock, and acid treatment as well as for growth in rich media (YPD) [39].

To determine whether the shared *HSP104* and *HSP78* motifs are in general use or specific to the *clpB* genes, we conducted whole-genome searches for each sequence. A search of the *S. cerevisiae* genome using an IUPAC consensus of the extended dot plot motif DP-I revealed that the *HSP78* and *HSP104* loci are the only two in *S. cerevisiae* with the DP-1 motif (Table 1, Model 1).

To identify highly similar variants, we implemented a position-weighted matrix (PWM) description of DP-I (Table 1, Model 2). This more permissive model identified 12 loci, half of which are related to a specific subset of proteins involved in various aspects of unfolded protein binding, disaggregation, and refolding. For example, one of the 12 loci corresponds to the promoter for *SSA1*, which encodes a member of the Hsp70/DnaK family of chaperones [40], which bind unfolded proteins to promote proper re-folding and prevent aggregation. In yeast, Ssa1 has known roles in disassembling misfolded protein aggregates together with Hsp104 [41], and in localizing mRNAs to the outer mitochondrial membrane for co-translational translocation and import into the mitochondria [42]. Hsp70-family proteins also interact with Hsp78. In conjunction with the mtHsp70 chaperone Ssc1, Hsp78 resolubilizes proteins from insoluble aggregates within mitochondria. Ssc1 is also involved in mitochondrial inner membrane import and refolding.

To improve our first PWM model, we constructed a second PWM descriptor based on the HSE4s upstream of the two *clpB* orthologs as well as *SSA1* (Table 1, Model 3). With this modified PWM-based model, fewer matches in ORFs, which are likely spurious, and an additional match in the promoter for the ubiquitin-encoding locus, *UBI4*, were identified. Ubiquitin is conjugated to misfolded proteins, marking them for proteolysis rather than refolding. Thus, this PWM descriptor improves the precision of the model with 7 loci of 9 involved in the disaggregation of misfolded proteins, their refolding, or their proteolysis. Sequence alignments of these targets indicate that the *clpB*-type HSE4 defines a set of four inverted repeats with the sequence 5′-**nGAAn**, which can also be described as two pairs of the palindrome 5′-**TTCyrGAA** separated by 2 bp (IUPAC DNA code: **y** = **C** or **T**, and **r** = **A** or **G**, **n** = any letter).

To test the significance of the central "**YR**" in the HSE4 motif, we queried the genome with an HSE4 motif in which these residues are not specified (Table 1, Model 4). We identified only 12 loci, including one new proteostasis gene, *HSP42*, which encodes a small heat shock protein that plays an important role in aggregate sorting and mediating the retention of misfolded proteins including amyloids in a peripheral cellular compartment [43,44]. The idealized (perfectly palindromic) HSE4 element, 5′-**TTCyrGAAnnTTCyrGAA**, is found at only six loci (*SSA1*, *HSP78*, *BTN2*, *UBI4*, *HSP104*, *SIS1*) in the entire genome, all of which are located at promoters of genes involved in protein homeostasis, including both *clpB* orthologs (Table 1, Model 5). *SIS1*, like *APJ1* (Table 1, Model 3) encodes a J-domain (DnaJ) family member that functions as co-chaperones to specific Hsp70 (DnaK) chaperones [45,46]. The *BTN2* gene encodes a sorting factor that accumulates in stressed cells to promote the transit of misfolded proteins to specific protein deposition sites, which are determined by alternative complex formation between Btn2 and Hsp42 (Table 1, Model 4) or Sis1 [44,47,48]. The results of these analyses are consistent with the hypothesis that yeast *HSP78* and *HSP104* are co-expressed together with a small set of other protein folding genes. This suggests that the two ClpB chaperones may participate in a single process related to protein homeostasis.

To investigate whether yeast *HSP78* and *HSP104* are indeed co-regulated, we used the SPELL (Serial Pattern of Expression Levels Locator) algorithm and query engine to survey their transcriptional correlation across 9,190 gene expression microarray experiments [31]. We found that *HSP78* was most significantly co-expressed with *HSP104* and *HSP42* (red arrow in Fig. 4D left panel). Similarly, *HSP104* was most significantly co-expressed with *HSP42* and *HSP78* (red arrow in Fig. 4D center panel). Furthermore, *HSP78* alone, *HSP104* alone, and *HSP78* plus *HSP104* in combination are co-regulated with diverse J-domain proteins and Hsp70/DnaK co-chaperones, which are known to stabilize unfolded protein states, respectively (J and K in Fig. 4D). This Hsp70 chaperone machinery is also co-expressed with other HSE4-associated genes such as *HSP42*, *BTN2*, *UBI4*, supporting the functional reliability of the HSE4 signature (Fig. 4D).

**Fig. 4. Yeast *HSP78* and *HSP104* share a unique regulatory architecture.**

(**A**) A dot plot between 320 bp sequence blocks located -10 bp and -65 bp upstream of the *S. cerevisiae HSP78* (vertical) and *HSP104* (horizontal) open reading frames, respectively. These coordinates were chosen because it anchors their HSE4 elements at the same position (starting 80 bp from the left) for ease of comparison. This comparison reveals a serial ordered alignment corresponding to an HSE4 element (blue box), a short alignment partially overlapping a triple array of STRE elements (pink boxes), and a TATA-containing core promoter element (green box).

(**B**) Graph of the relevant details of the *HSP78* (top) and *HSP104* (bottom) upstream regulatory sequences. Each track (numbered lines 1–6) provides locations of the dot plot motifs (DP-I and DP-II) and known binding motifs (colored boxes in lines 1–5). Labels above certain binding sites indicate binding of a known transcription factor at this position under the specific conditions listed [39,49]. Line 6 shows conserved non-coding sequences (CNS) as determined from comparisons of corresponding orthologs from other species of *Saccharomyces*. Dots and lines in the ruler mark off 10 bp and 100 bp intervals, respectively. The transcriptional arrows represent only approximate positions of the transcriptional +1 based on distance from the TATA-containing promoter motif.

(**C**) Motif key for panel B.

(**D**) Results using the Serial Pattern of Expression Levels Locator (SPELL [31]) algorithm/query engine to identify the closest regulated genes to *HSP78* (left graph), *HSP104* (center graph), or both genes combined (right graph). Plotted are the transcriptional correlation scores (SPELL, Adjusted Correlation Scores). The SPELL scores are based on pair-wise Pearson correlation co-efficients with Fisher z-transforms across 9,190 gene expression studies in yeast [31].These graphs show that the two *clpB* genes are co-regulated with one another (red arrows) and to Hsp70/DnaK (K) and J-domain/DnaJ co-chaperones (J), which bind and recruit unfolded proteins to chaperone machinery in the cytoplasm (blue bars) and in the mitochondria (purple bars).

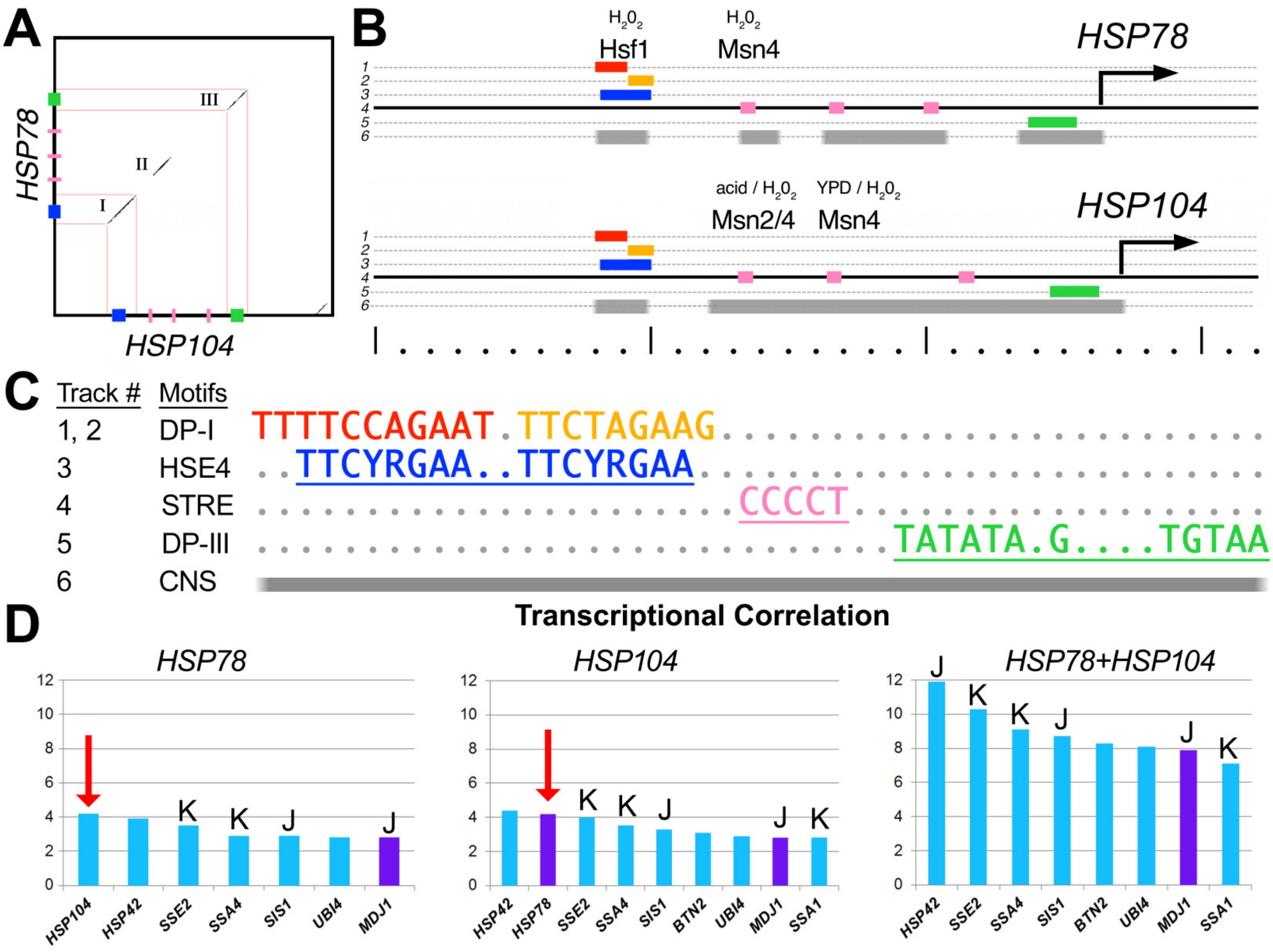

## HSE4 regulation of the *clpB* genes is conserved in choanoflagellates.

To determine whether HSE4 regulation is also present in the choanoflagellates we looked for HSE4 signatures in the *M. brevicollis* and *S. rosetta* genomes, which maintain the *clpB* genes. Our analysis of the *S. rosetta* genes revealed the presence of a consensus HSE4 sequence upstream of the *HSP78* gene and a near consensus HSE4 sequence upstream of *HSP104*. We identified a total of 5 additional HSE4 sequences in genome-wide searches for a consensus HSE4 (Table 2). One HSE4 is located upstream of *UBA1*. Three additional HSE4s are found upstream of unrelated genes, and one lies within an open reading frame. *UBA1* encodes a ubiquitin-like modifier activating enzyme, which catalyzes the first step in ubiquitin conjugation, marking proteins for degradation. HSE4 regulation of *UBA1* in the choanoflagellate is reminiscent of HSE4 regulation of ubiquitin (*UBI4*) in *S. cerevisiae*.

We also find more extended Hsf1 binding arrays, here "HSE4+" elements, at the *clpB* loci of the choanoflagellate *M. brevicollis* and chytrid fungus *B. dendrobatidis*. These match so-called extended "skipped HSE" motifs [32] in which the pattern consists of five or six inverted 5′-**nGAAn** repeats (HSE5, HSE6) with one or more missing internal triplet **TTC**'s or **GAA**'s. For these genomes, we can derive lineage-specific consensus signatures that are specific to a small *clpB* regulon. For example, in *M. brevicollis* we find a consensus skipped HSE at exactly 6 loci that include *HSP104*, *HSP78*, and an Hsp70 gene, while in the chytrid *B. dendrobatidis* we find only five loci, including *HSP78* and *HSP104*, with either an HSE4 (5′-**TTC..GAA..TTC..GAA**) or a skipped HSE5 (5′-**TTC..GAA..TTC..G....TTC**). Last, we note that even in the yeast *S. cerevisiae* there is a single locus matching a skipped HSE5 motif, although in this case it is a perfect HSE5 (*i.e.*, not skipped) and is associated with the *SIS1* DnaJ-encoding locus that we already identified as having an HSE4 motif. Overall, these results show that the archetypical HSE4/HSE4+ elements are ancestral to a small regulon that includes *HSP78* and *HSP104* and that these genes evolve with only modest changes to the Hsf1 binding site motif.

## Eukaryotic ClpB promoter architecture is unique and shared.

To identify the extent to which the *clpB* promoter architecture is present in the *S. cerevisiae* genome, we searched for the combination of the HSE4 motif and the STRE cluster. We relaxed the stringency of the combined motifs until we found the closest matching sequences. For example, a search for all 160 bp windows containing at least three STREs (5′-**CCCCT**) and a match to either a skipped (5′-**TTCnnGAAnnnnnnnGAA**) (Table 1, Model 6) or an end-clipped HSE4 (5′-**CnnGAAnnTTCnnG**) identifies the same three loci: *HSP42*, *HSP78*, and *HSP104*. The spacing of the triplet cluster at *HSP42* is not similar to that found at *clpB* promoters and the STRE elements at *HSP42* are not conserved across the *Saccharomyces* genus [49], nor is the *HSP42* family found in choanoflagellates.

Thus, yeast *HSP78* and *HSP104* are uniquely associated with a shared promoter architecture, one element of which (HSE4s and skipped HSE5/6 arrays) is still uniquely associated with their orthologous promoters in choanoflagellate genomes.

## Absence of Hsp78 exacerbates hsp104 mutant thermosensitivity.

The co-regulation and co-expression of both mitochondrial and cytoplasmic ClpB functions is consistent with a potential joint function. To experimentally assess this hypothesis, we tested in *S. cerevisiae* whether a phenotypically normal *hsp78* deletion had any effect on *hsp104* mutant thermosensitivity. In these experiments, single and double mutants were subjected to extreme heat (50$^{o}$C) in rich media (YPD) following pre-exposure to mild heat sufficient to induce *HSP104* and *HSP78* gene expression. As expected, survival of the *hsp78* deletion mutant was indistinguishable from that of wild type ($p = 0.9$) and viability of the *hsp104* mutant was reduced relative to that of wild type strains (Fig. 5A). This is consistent with previous reports showing that *hsp104* deletion strains are measurably thermosensitive [50-52] while *hsp78* deletion strains have more subtle effects when heat stressed [24-26,53-55]. However, we found that the *hsp78 hsp104* double mutant was significantly more sensitive than *hsp104* single mutant (~10-fold difference, $p = 0.0005$) suggesting that Hsp78 contributes to induced thermotolerance in the absence of Hsp104, although it has no appreciable effect in its presence (Fig. 5A). This revealed for the first time that survival in extreme heat depends on the presence of both compartmentalized ClpB functions. Together with the observation of co-loss and co-regulation of *clpB* orthologs, this result leads us to postulate a shared ClpB function.

**Fig. 5. Hsp78 and Hsp104 are required for thermotolerance.**

Viability of *clpB* mutants cultured in rich media (YPD). Cultures were grown to log phase at 30℃, pre-treated at 37℃ and diluted 1:1 into 50℃ media (20 min) or diluted directly into cold water. Values are average colony forming unit (cfu) counts from YPD plates incubated at 30℃ from three strains of each genotype. Error bars correspond to the standard deviation. Strains are: wild type: BY4730, BY4700, BY4738; *hsp78*: JF2478, JF2479, JF2480; *hsp104*: JF2473, JF2474, JF2498; *hsp78 hsp104*: JF2494, JF2495, JF2516. Double asterisks correspond to a p-value of $< 0.001$ in an unpaired Student's *t*-test.

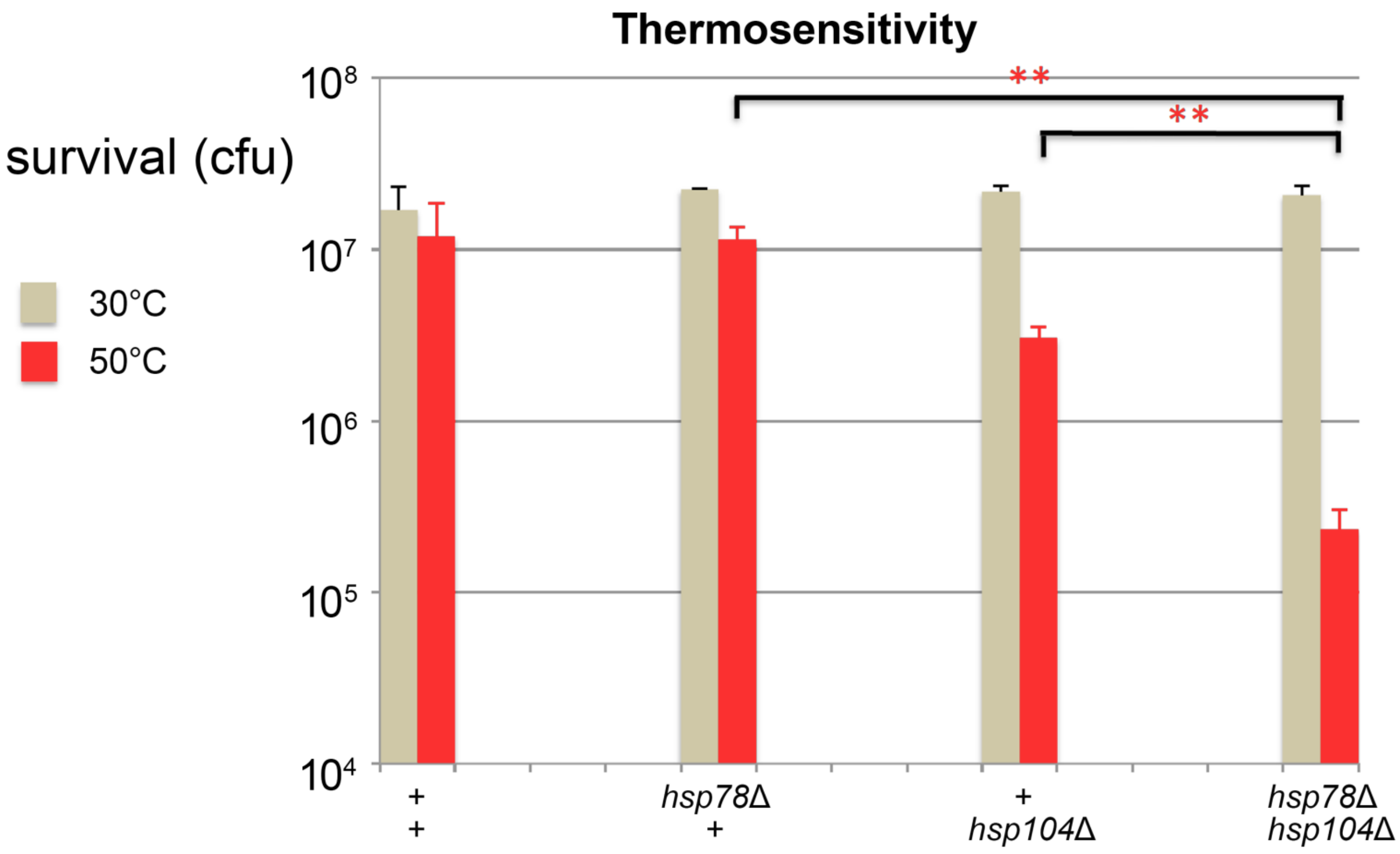

## Yeast interactome indicates that the ClpB proteins share specific protein clients.

Connections between the lost genetic functions can be inferred from gene association algorithms (e.g., GeneMania [16,17]) that identify physical or genetic interactions among the lost genes. We identified several links between genes in the different biosynthetic pathways (blue nodes in Fig. 6A) and other lost genes that may be informative upon further study. Here, we focus on several physical links connecting both of the ClpB chaperones (red nodes in Fig. 6A) with Lys4 and Lys20. Lys20 and Lys4 occupy early steps in the $\alpha$-aminoadipate lysine biosynthesis pathway, which is found in a small group of thermophilic archaea and bacteria and in non-metazoan eukaryotes [56-58]. Lys20 is the mitochondrial isoform of homocitrate synthase that exerts most of the flux control in lysine biosynthesis in yeast [59], while Lys4 encodes mitochondrial homoaconitase, a 4Fe-4S iron-sulfur protein [60].

The observation that Hsp78 and Hsp104 each make specific contacts with Lys4 and Lys20 (Fig. 6A) despite their compartmentalized roles, leads us to propose that *clpB*-encoded chaperones Hsp104 and Hsp78 were lost because of "lost *shared* clients" that included Lys4 and Lys20, but may also have included other nuclear-encoded, cytoplasmically-translated, and mitochondrially imported proteins. This idea is investigated further below.

## Lys4 and Lys20 interact specifically with RAC and ClpB chaperone systems.

If Lys4 and Lys20 are heat-labile clients specifically disaggregated by Hsp78 and Hsp104 after heat denaturation, they are likely also to require folding assistance *de novo* during translation and/or co-translationally with mitochondrial import. We examined whether Lys4 and Lys20 make specific contacts with the ribosome-associated chaperone (RAC), which is important for de novo folding during translation and leads to aggregation when compromised [61-63]. The RAC/Ssb system is also required during co-translational translocation into subcellular compartments [64]. RAC is composed of the Hsp70 family chaperone Ssz1 tightly associated with its J-domain co-chaperone Zuo1 and more loosely with a second Hsp70 family member, Ssb (Ssb1 or Ssb2). To investigate the role of RAC in Lys4 and Lys20 folding, we used experimental interaction data to construct the interactome among the Lys4 and Lys20 proteins, the RAC system and various Hsp70/DnaK and J-domain/DnaJ proteins that we identified in our gene regulatory analyses (Fig. 6B).

These analyses show that Lys4 and Lys20 interact with RAC machinery and the ClpBs but not with other Hsp70 chaperones and DnaJ co-chaperones. For example, Lys4 interacts physically with both components of RAC (the Hsp70 Ssz1 and its dedicated J-domain co-chaperone Zuo1), and with the ribosomal Hsp70s Ssb1 and Ssb2, which work in conjunction with RAC (Fig. 6B). But Lys4 does not interact with diverse other cytoplasmic (Sse2, Ssa4, Ssa1) or mitochondrial (Ssc1) Hsp70s nor with J-domain family members (Sis1, Ydj1, Apj1, Mdj1), several of which (*SSE2*, *SSA4*, *SSA1*, *SIS1*, MDJ1) are co-expressed with the *clpB*s (Fig. 4D), or definitively co-regulated with the *clpB*s (*SSA1*, *SIS1*, *APJ1*) (Table 1) based on the presence of HSE4 elements. A similar result was seen for Lys20. We thus propose that unfolded Lys20 and Lys4 are recognized by RAC during translation as nascent unfolded polypeptides and again by ClpBs after post-translational heat denaturation.

**Fig. 6. Specific interactions between the lost lysine biosynthetic enzymes and the lost ClpB chaperones in yeast.**

(**A**) Yeast interactome graph showing high-throughput physical interactions (pink links), medium throughput TAP-tag physical interaction data for diverse chaperones (gray links), and genetic interactions (green links) amongst the yeast counterparts to the twenty-four lost genetic functions. Genetic interactions are based on 108 separate data sets, while physical interactions, not including the separate TAP-tag data set, are based on 103 data sets. The gene nodes are also colored by functional category. The largest statistically significant category of enrichment is for "biosynthetic process" with an expect value in random sampling of 2.3E-24 (17/27 lost genes). The two lost eukaryotic ClpB chaperones encoded by *HSP104* and *HSP78* interact physically with only two of the lost genes, *LYS4* and *LYS20*, which are mitochondrial components of the lysine biosynthesis pathway.

(**B**) Gene association analysis shows that Lys4 and Lys20 make contact with ribosome-associated chaperones (RAC) but no other Hsp70 and J-domain co-chaperones. Proteins localized to the mitochondria are indicated with purple labels. Genes with known roles in protein targeting and import to the mitochondria or refolding after import are indicated by asterisks (see key). Genes with HSE4 elements in their promoters (Table 1) are indicated by yellow halos. The edge and node color combination is as in (A).

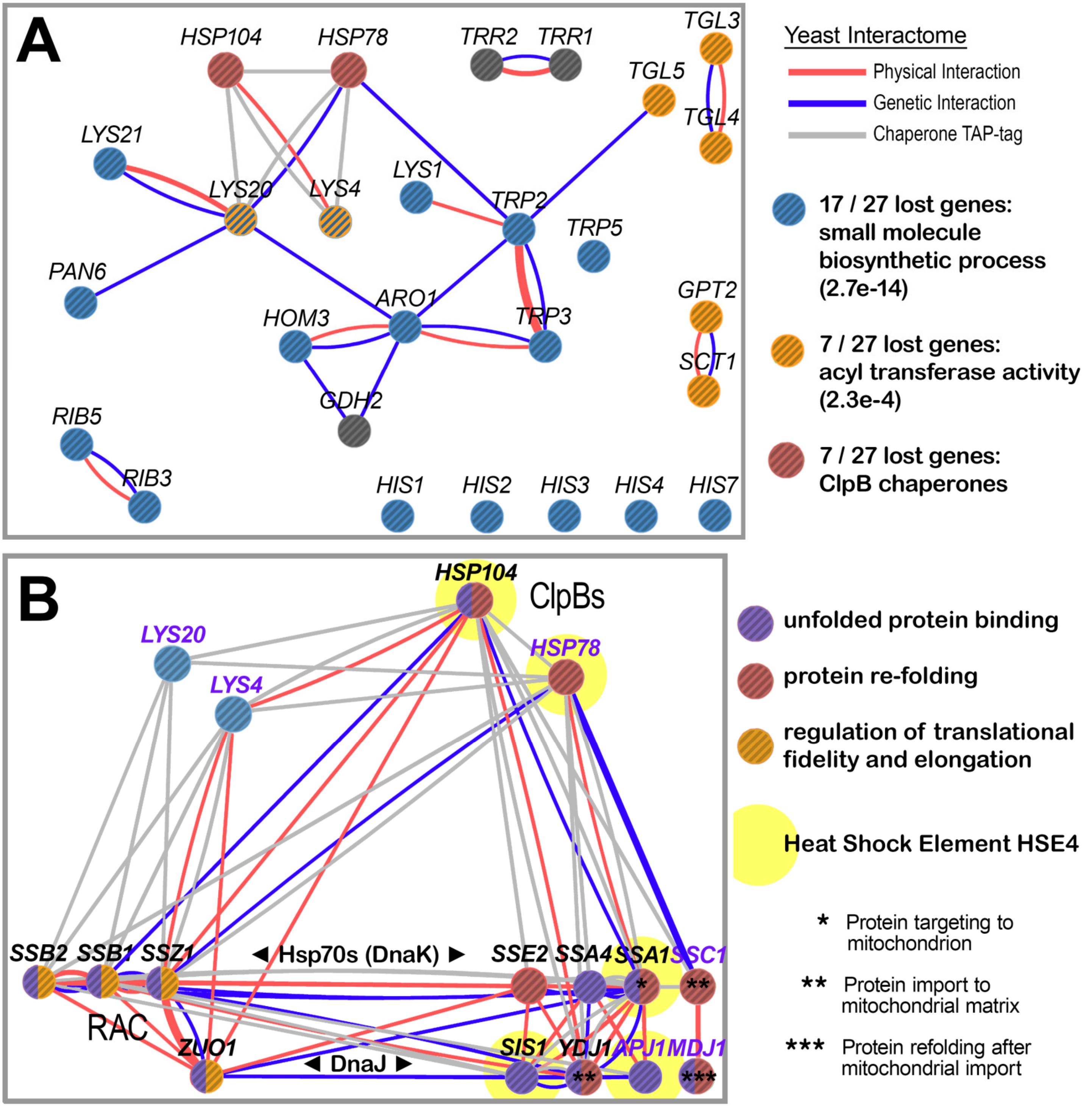
A
HSP104
HSP78
TRR2
TRR1
TGL3
TGL5
TGL4
LYS21
LYS20
LYS4
LYS1
TRP2
TRP5
PAN6
ARO1
HOM3
TRP3
GPT2
SCT1
GDH2
RIB5
RIB3
HIS1
HIS2
HIS3
HIS4
HIS7
Yeast Interactome
Physical Interaction
Genetic Interaction
Chaperone TAP-tag
17 / 27 lost genes: small molecule biosynthetic process (2.7e-14)
7 / 27 lost genes: acyl transferase activity (2.3e-4)
7 / 27 lost genes: ClpB chaperones
B
HSP104
ClpBs
LYS20
HSP78
LYS4
SSB2
SSB1
SSZ1
◄ Hsp70s (DnaK) ►
SSE2
SSA4
SSA1
SSC1
RAC
ZUO1
◄ DnaJ ►
SIS1
YDJ1
APJ1
MDJ1
unfolded protein binding
protein re-folding
regulation of translational fidelity and elongation
Heat Shock Element HSE4
* Protein targeting to mitochondrion
** Protein import to mitochondrial matrix
*** Protein refolding after mitochondrial import

**Fig. 7. Several additional Fe/S proteins related to Lys4 are also missing in holozoans.**

(**A**) Lys4 carries an [4Fe-4S]-type cluster, which is essential to its enzymatic functions and is a potential generator of reactive oxygen species (ROS) [68,69].

(**B**) Listed are *LYS4* and four additional Fe/S protein-encoding genes related to *LYS4*. Four out of five of these (*LYS4*, *ACO2*, *ILV3*, and *LEU1*) are retained (+) in fungi and the unicellular choanoflagellate *M. brevicollis* (*M. bre.*), partially retained in the colonial choanoflagellate *S. rosetta* (*S .ros.*), and entirely undetectable (-) in metazoa. These genes are involved in amino acid biosynthesis (leucine, lysine and branch chain amino acids (BCAA) except for *ACO1*, which is the cytoplasmic aconitase of the TCA/citric acid cycle.

**(C)** An annotated sequence alignment for Lys4, Aco2, and Aco1 peptide sequences around the Fe/S coordination domain, which is characterized by three cysteine residues conserved in the Lys4 superfamily that includes Aco1/2. The alignment is annotated to show the family-specific residues of Lys4 (cyan shading), Aco2 (green shading), and Aco1 or Aco1/Aco2 (gray shading). Hydrophobic amino acids are indicated by bold red letters. Small neutral or hydrophobic amino acids are indicated by small letters: Gly (g), Ala (a), Ser (s), and Pro (p). Cysteines (C) involved in Fe/S coordination are underlined and highlighted in yellow. Family-specific amino acids that are hydrophobic are indicated by shaded asterisks for Lys4 (top) and Aco1 (bottom).

**(D)** Hydrophobic amino acids constitute 41.4% of Lys4 (12/29) and 22.8% of Aco1 (8/35) family-specific residues, respectively. Gene names are based on names of orthologous *S. cerevisiae* genes. Aco1 sequences: XP_001747281.1, *Monosiga brevicollis* (*Mb*); XP_004989369.1, *Salpingoeca rosetta* (*Sr*); and AAH26196.1, *Homo sapiens* (*Hs*). Aco2 sequences: NP_012335.1 (*Sc*) and XP_001746450.1 (*Mb*).

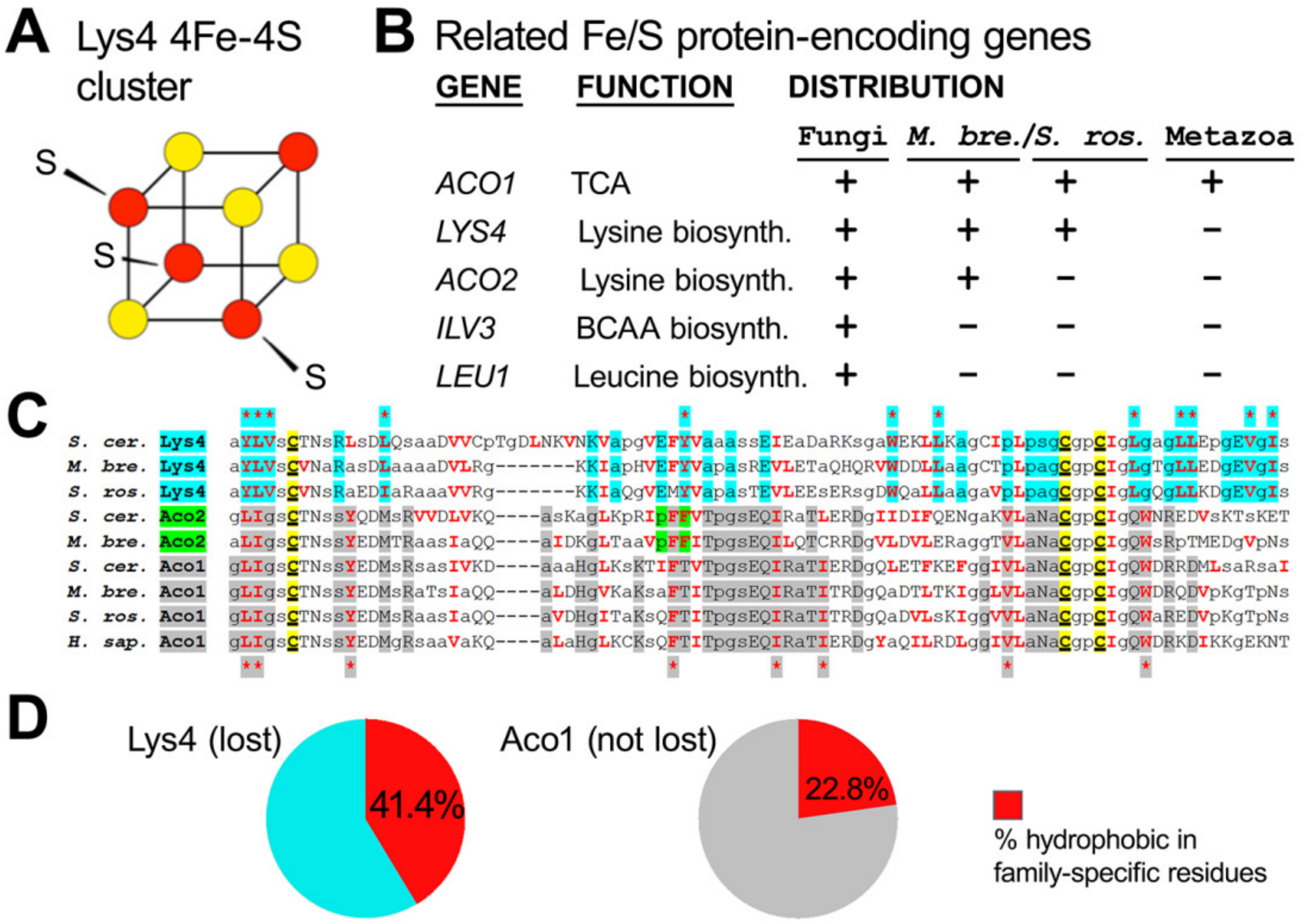
A Lys4 4Fe-4S cluster
S
S
S
B Related Fe/S protein-encoding genes
GENE FUNCTION DISTRIBUTION
Fungi M. bre./S. ros. Metazoa
ACO1 TCA + + + +
LYS4 Lysine biosynth. + + + −
ACO2 Lysine biosynth. + + − −
ILV3 BCAA biosynth. + − − −
LEU1 Leucine biosynth. + − − −
C
S. cer. Lys4
M. bre. Lys4
S. ros. Lys4
S. cer. Aco2
M. bre. Aco2
S. cer. Aco1
M. bre. Aco1
S. ros. Aco1
H. sap. Aco1
D
Lys4 (lost)
41.4%
Aco1 (not lost)
22.8%
% hydrophobic in family-specific residues

## Other Fe/S proteins in same superfamily are lost in choanoflagellates and animals.

If Lys4 and Lys20 were lost in animals due to their deleterious misfolding properties, then other proteins with similar structures may have had similar fates. Indeed, additional Lys4 superfamily proteins containing 4Fe-4S clusters (Fig. 7A) were also lost in the evolution of animals. We looked for examples in which other Lys4 superfamily members were lost in choanoflagellates and had therefore escaped our initial screen for losses in the stem-metazoan lineage. While Lys4 is absent in animals but present in choanoflagellates (Table 1 and Table S1), we found that *ACO2* is absent in animals and the colonial choanoflagellate *S. rosetta*, but present in the unicellular choanoflagellate, *M. brevicollis*, and that *LEU1 and ILV3* are absent in both animals and choanoflagellates (Fig. 7B). Like Lys4, both Ilv3 and Aco2 receive their Fe/S clusters in the mitochondria [60] although Leu1 is a cytoplasmic enzyme [58]. Fe/S proteins have two special properties; first, the free iron that is presumably released upon stress-mediated denaturation can generate hydroxyl radicals, which are one type of insidious reactive oxygen species (ROS) [65,66]. Second, Fe/S cluster proteins are highly susceptible to inactivation by superoxide, which accumulates and contributes to the loss of viability in heat stressed cells [67-70].Thus, Lys4, which is among the most highly conserved proteins in our list of twenty-four lost genetic functions (Fig. 1, Table S1), may be especially heat labile and prone to ROS generation when misfolded in the presence of oxygen. The same may be true of other Lys4 relatives. We therefore speculate that this class of protein may require dedicated folding machinery.

Hydrophobic amino acids, which are normally packed into the interior of a peptide, are recognized by yeast Hsp104 when exposed in unfolded aggregates and when added to model clients increase their recognition after denaturation [71]. We therefore compared the hydrophobicity of the Fe/S coordination pocket of Lys4 with that of Aco1, which is maintained in animals. We first aligned the yeast and choanoflagellate peptide sequences from these proteins, including those from Aco2 and annotated the family-specific residues for Lys4 (cyan shading in Fig. 7C) and Aco1 (gray shading). We found that 41.4% (12/29) of the conserved Lys4-specific residues are hydrophobic compared with 22.8% of the conserved Aco1-specific residues (Fig. 7D). As the hydrophobic content of a protein is likely to be positively correlated with aggregation after heat denaturation, the two-fold enrichment in Lys4 relative to Aco1 is likely to pose a significant cellular challenge.

# Discussion

## Hsp78 and Hsp104 chaperones function coordinately and may have common clients in Lys4 and Lys20.

Here we show for the first time that Hsp78 and Hsp104 (***i***) were both definitively lost in an ancestor of animals despite their presence in all non-animals; (***ii***) were uniquely co-regulated by a multimeric and ordered array of perfect binding sites for Heat Shock Factor 1 (Hsf1) in an ancestor of animals based on the shared presence and restricted use of this regulatory signature in fungal and choanoflagellate genomes; (***iii***) function together in thermotolerance (Fig. 5A); and (***iv***) make specific contacts with Lys4 and Lys20, two nuclear encoded, mitochondrial enzymes (Fig. 6A and 6B).

These results are consistent with the idea that specific mitochondrial biosynthetic enzymes are shared clients of both ClpBs and could be pertinent to their joint loss at the base of animals. We propose that cytoplasmic Hsp104 disaggregates and refolds nuclear-encoded, cytoplasmically translated mitochondrial enzymes such as Lys4 and Lys20 that are also re-folded by Hsp78 in the mitochondria. However, exactly how compartmentalized Hsp104 and Hsp78 chaperones could share mitochondrial clients is unclear. One idea is that there are parallel co-translational import and mitochondrial targeting pathways for getting Lys4 into the mitochondrial matrix. In this case, Lys4 apoprotein *en route* to the mitochondria might exist as a cytoplasmic pool that is prone to misfolding and aggregation. A second idea is that thermally denatured, incompletely translocated Lys4 enzymes might escape into the cytoplasm, which would create a need for maintaining a cytoplasmic ClpB activity. Finally, a third possibility is that Lys4 is able to escape via compromised mitochondrial membranes, which are thermosensitive. For example, even mild heat (37°C), sufficient to induce the ClpB chaperones, results in a 5-fold increase in mitochondrial leakage based on the transfer of mitochondrial DNA to the nucleus [72].

Despite our findings of a possible Lys4-ClpB client-chaperone relationship, it is not likely that Lys4 was the only client of the ClpBs in the stem-metazoan lineage. First, yeast data has implicated several clients of Hsp104 [73]. Second, ClpBs are conserved in plants and bacteria, but Lys4 is not, because plants and most bacteria utilize a lysine biosynthesis pathway different from the one in opisthokonts [56]. For these reasons, we suspect that the ClpBs in animals were lost not so much because a single dedicated client was lost, but because a threshold number of clients were lost.

## A cascade of gene loss in the stem-metazoan lineage.

Our results suggest that many metabolic genes were lost in the stem-metazoan lineage (Fig. 1). This may reflect wholesale reduction of biosynthetic capacity in favor of obligate high-throughput filter feeding and could be due to the trophic benefits of multicellular organization, allowing biosynthetic precursors to be derived as nutrients from dietary sources. For example, sponges, which have diverse complex multicellular chambers for pumping and filtering water, are known to take up and remove up to 99% of bacteria and 94% of small eukaryotes from the water that they filter [74,75]. Thus, we propose that the stem-metazoan innovation resulting in efficient dietary intake of picoplankton via grazing and/or filtering is likely to have resulted in loss of selection of certain biosynthetic pathways because it became more energetically efficient to obtain them from feeding rather than from biosynthesis. We suggest that this working hypothesis may lead to deeper insights into the evolutionary emergence of specialized-feeding surfaces and the forerunners of the "gastraea" [76].

Based on our finding of specific gene losses, we propose a two-step model in which loss of biosynthetic enzymes (step 1) is followed by secondary loss of attendant chaperones (step 2). This model is supported by several additional observations in choanoflagellates. First, we note that the *TRP2* gene, which encodes the first enzyme in the energetically-costly tryptophan biosynthetic pathway is not as well conserved in choanoflagellates as the intermediate enzyme encoded by *TRP3*, and is more poorly conserved than *TRP5*, which encodes the final enzyme in tryptophan synthesis (Fig. 1 and Table S1). Thus the extent of conservation is co-linear with position along the biosynthetic pathway, consistent with the possibility that choanocyte-type filter feeding [77] and possibly facultative coloniality in this lineage led to reduced selection of the early components of this pathway while the final biosynthetic enzymes were maintained for the ability to complete tryptophan biosynthesis using intermediate precursors provided by ingested microorganisms. Second, in the course of testing various computational approaches to gene loss, we discovered that some "lost" animal genes are present in the unicellular choanoflagellate *M. brevicollis* but not in the facultatively colonial [12] choanoflagellate *S. rosetta* or in animals (*e.g.*, *ACO2* in Fig. 3B). Third, we did not find any of the genes encoding enzymes in the biosynthetic pathways for the branched chain amino acids, (leucine, isoleucine, and valine), in either animals or choanoflagellates. These amino acids are three of the ten essential amino acids animals must retrieve from their dietary intake. The pattern of metabolic gene loss in choanoflagellates suggests that (*i*) specific metabolic gene losses had already occurred in the stem-holozoan ancestor or were underway with holozoan divergence; (*ii*) additional metabolic gene losses were sustained during the divergence of choanoflagellates and animals (*e.g.*, *ACO2*); (*iii*) gene losses in choanoflagellates began with genes encoding enzymes at the beginning of pathways (*e.g.*, Trp enzymes); and (*iv*) metabolic genes were more completely lost with the evolution of animals. These results are likely to prove fundamental for understanding formative events in the evolution of the metabolic physiology of the animal kingdom.

# Materials and Methods

**Comparative genomics.** Ensembl's BioMart tool [21] was used iteratively to query, sort, and identify orthology calls [78,79] for Ensembl Genes Release 75 (www.ensembl.org) and Ensembl Fungi Genes Release 22 (fungi.ensembl.org). The choanoflagellate genome of *Monosiga brevicollis* (v1.0) was queried from the Joint Genome Institute databases (genome.jgi.doe.gov), the *Salpingoeca rosetta* genome from Broad Institute databases (www.broadinstitute.org/). Note that BioMart orthology calls are specific to a release set of genome comparisons and that orthology calls sometimes change with each Release. All genomes were also queried using taxonomic specification at NCBI's BLASTP server, except for the ctenophore *Mnemiopsis leidyi*, which is available via National Human Genome Research Institute (NHGRI) server (research.nhgri.nih.gov/mnemiopsis). Homologs for fungal lineages outside of Ascomycetes and other non-fungal lineages were selected following BLASTP analysis of specific genomes using *S. cerevisiae* proteins as queries. Additional filtering using E-value thresholds is described in the text. Data from a Boolean-type PFAM taxonomy search (*Dictyostelium* protein domains absent in yeast and animals) was conducted at http://pfam.xfam.org. The identified PFAM domain IDs were then used to retrieve the protein sequences using BioMart [15,21]. These sequences were used in a BLASTP query of choanoflagellates. Orthologs of lost genes from *S. cerevisiae* or *D. discoideum* are listed in File S1.

**Phylogenetic analyses.** Alignments were produced using ClustalW. Bayesian Markov Chain Monte Carlo analyses were conducted using MrBayes [80-82] and initial runs sampled diverse fixed amino acid rate matrices, but most data sets overwhelmingly supported the WAG [83] (ClpB trees) or Jones/JTT [84] (Lys4 trees) models after the 25% burn-in phase. Topological convergence was defined as having achieved an average standard deviation of split frequencies $< 0.01$, but we typically achieved $< 0.002$. Maximum Likelihood and exploratory Neighbor-Joining trees were conducted in MEGA5 and MEGA6 [85,86] typically using Jones/JTT matrices and a gamma distributions with shape parameters spanning 0.7 to 1.0 with replicate boot strap support based on 500 replicates. Alignment files and supporting Bayesian analysis files are archived in File S2.

**Gene Association Analysis.** Gene association data pertaining to individual chaperones was retrieved from ChaperoneDB (chaperonedb.ccbr.utoronto.ca) and uploaded to GeneMania [16,17] (www.genemania.org). Evidence for physical interactions is represented by red edges, evidence for genetic interactions is shown by green edges and predictions based on selected data imported from the global TAP tag MS dataset [73] is shown by gray edges.

**Regulatory Bioinformatics.** Identification of matching cis-regulatory patterns in whole-genome sequences were conducted by downloading text files of each respective genome, removing newlines, and searching for blocks of sequence containing the indicated patterns (Tables 1 and 2). These searches were conducted using regular expression patterning matching in simple UNIX shell scripts incorporating perl, grep, and standard shell commands. BLASTN-based dot plot analysis was conducted using a window size of 320 bp for sequences upstream of *S. cerevisiae HSP78* (vertical) and *HSP104* (horizontal) open reading frames. Known transcription factor binding sites at listed positions were retrieved using the UCSC genome browser [39,49]. Conserved non-coding sequences (CNSs) were determined from comparisons of corresponding orthologs. Transcriptional correlation scores correspond to the Adjusted Correlation Scores for 430 datasets representing 9190 total arrays from 286 published studies queried via SPELL [31] (Serial Pattern of Expression Levels Locator) and available through the *Saccharomyces* Genome Database (spell.yeastgenome.org).

**Yeast Strains and Survival Assay.** Yeast strains were constructed by mating and tetrad dissection of diploids starting with strains from the MAT$\alpha$ haploid deletion collection (BY4742). Deletion collection strains were mated with BY4730 and diploids sporulated and dissected to generate *LYS2*+ derivatives of the *hsp78* and *hsp104* mutants (*hsp78*: JF2478, JF2479, JF2480 and *hsp104*: JF2473, JF2474 and JF2498). These strains were crossed to generate the double *hsp78 hsp104* mutants (JF2494, JF2495, JF2516). Strain genotypes are provided in Table S2 in File S3. *clpB* genotypes were confirmed by colony PCR using a forward primer upstream of the deletion and a reverse primers within the kanMX cassette. Primer sequences are provided in Table S3 in File S3.

Survival was assayed by colony forming unit (cfu) assays. Cultures were grown to saturation in YPD at 30°C, then subcultured and grown to log phase under the same conditions. Log phase cultures were shifted to 37°C for between 40–60 minutes and then diluted (1:1) into ice-cold water or diluted similarly into 50°C YPD media for 20 minutes. Cultures were then serially diluted and plated on YPD. Colony number was used to calculate the concentration of survivors for each strain and each treatment condition. Values for individual strains were averaged. SD, standard deviation.

## Acknowledgements and funding information

We thank J. Logsdon, R. A. Cornell, D. Weeks, G. Gussin and B. Phillips for feedback on versions of this manuscript.

## Supporting Information available in 2015 *PLoS ONE* version (see link below)

Table S1. List of twenty-four lost genetic functions

File S1. FASTA file of list of 28 genes from Fig. 1

File S2. Zipped archive with all NEXUS alignment files and associated Bayesian analyses.

File S3: Tables S2 and S3:

Table S2. Yeast strains used in this study

Table S3. Primer sequences used in this study